\documentclass[12pt]{article}
\usepackage{amsfonts}
\usepackage{epsf}
\def\CellGroup{\bgroup}
\def\endCellGroup{\egroup}
\begin{document}
\thispagestyle{empty}
\vspace*{3cm}
\large
\begin{center}
{\Large\bf $T$-odd correlation in  the $K^+ \to \pi l \nu \gamma$ decays
beyond Standard Model }
\end{center}

\large
\vspace*{1cm}
\begin{center}
V.V. Braguta, A.A. Likhoded, A.E. Chalov 
\end{center}

\vspace*{1cm}
\begin{center}
Institute for High Energy Physics,  Protvino, 142280 Russia
\end{center}

\normalsize
\vspace*{2cm}
\underline{\bf Abstract}

\large
\vspace*{0.5cm}
\noindent

\noindent
The dependence of the $T$-odd correlation on the effective Lagrangian parameters in the $K^+ \to \pi l \nu  \gamma,~ l=e, \mu$ decays is analyzed. It is shown that 
the introduced observable is perspective in a search for new physics in the 
vector and pseudovector sector of the Lagrangian. As for scalar and pseudoscalar sectors, the $T$-odd correlation studies will not allow one
to improve current restrictions on parameters of models beyond SM.

\newpage
\large

\section{Introduction}

\noindent
$T$-invariance is one of the fundamental symmetries in physics, therefore 
many experimental groups carry out their studies in this area. The search for a new physics are extremely promising in the processes, where the Standard Model contribution to experimental observables is suppressed. 
One of such experimental observable, for instance, is the muon transverse polarization in $K^+ \to \pi^0 \mu \nu, K^+ \to \mu \nu  \gamma$ decays \cite{Abe:2002vc, Kudenko:2000sc}. In these processes the Standard Model contribution to $P_T$ vanishes at the tree level. Nonvanishing contribution appears only at one loop level and is caused by the electromagnetic final state interaction, therefore it is significantly suppressed.
In the $K^+ \to \pi^0 \mu \nu$ decay the lepton transverse polarization is equal to $5 \times 10^{-6}$\cite{Zhitnitsky:he, Efrosinin:2000yv}, and in $K^+ \to \mu \nu \gamma $ it is equal to $6 \times 10^{-4}$ \cite{Braguta:2002gz, Rogalev:2001wx}.

In contrast to SM, in some extensions the transverse polarization appears already at tree level of perturbation theory \cite{Belanger:1991vx, Chen:1997gf}.

At the moment E246 experiment at KEK performs the analysis of the data on 
the $K^+ \to \pi^0 \mu \nu$ process to put bounds on the $T$-violating muon transverse polarization, where the following result has been obtained \cite{Abe:2002vc}:

\begin{eqnarray}
P_T=(-1.12 \pm 2.17(stat.) \pm 0.90(syst.)) \times 10^{-3}.
\label{pt}
\end{eqnarray}

Unfortunately, there is no experimental result for the $K^+ \to \mu \nu \gamma$ process, as the the experimental data are still processed. The transverse polarization is expected to be 
of order  of $1.5 \times 10^{-2}$ \cite{Kudenko:2000sc}.

Another important experimental observable used in $CP$ - violation searches is the $T$-odd correlation in the 
$K^+ \to \pi^0 l \nu \gamma$ decay defined as $\xi = {\bf q} \cdot [{\bf p_{\pi}} \times {\bf p_l]} / {m_K^3}$.

In this case the $T$-violating signal is the asymmetry of differential distribution of 
decay width relative to the $\xi =0$ line. As in the case of lepton transverse polarization, $T$-odd correlation vanishes at tree level of SM and appears due to the electromagnetic final state interaction. In the framework of SM this effect was examined earlier in \cite{Braguta:2001nz}. However it would be interesting to compare that result with the value of the asymmnetry induced by some of SM extensions. Our work is devoted to this problem.

New perspectives on $T$-odd correlation researches are connected with OKA experiment 
\cite{Obraztsov:2000et}, where the measurement of this observable is planned.
For this reason, the problem under discussion is of particular importance.
The event samples of $10^6 \div 10^7$ for the 
$K^+ \to \pi^0 e \nu_{e} \gamma$ decay and $10^5 \div 10^6$ for $K^+ \to \pi^0 \mu \nu_{\mu} \gamma$ one are expected to be accumulated. 

This paper is organized as follows:
In section 2 we present the model independent Lagrangian, the expression for the assymetry in terms of the Lagrangian, discuss the SM contribution to $T$-correlation. In sections  3 and 4 the contributions of the $SU(2)_L \times SU(2)_R \times U(1)$ model and scalar models are examined. Last section summarizes the results and conclusions.

\section{Model independent approach in $T$-odd corelation study}

\noindent
The model independent Lagrangian of 4-fermion interaction is as follows:
\begin{eqnarray}
L= \frac {G_f} {\sqrt 2} \sin \theta_c \biggr ( \bar {s} \gamma^{\alpha} (1-\gamma_5) u 
\bar {\nu} \gamma_{\alpha} (1-\gamma_5) l +
g_s \bar s u \bar \nu (1+ \gamma_5) l +
\label {Lagr} \\ \nonumber
g_p \bar s \gamma_5 u \bar \nu (1+ \gamma_5) l +
g_v \bar s \gamma^{\alpha} u \bar \nu \gamma_{\alpha} (1 - \gamma_5) l+
g_a \bar s \gamma^{\alpha} \gamma_5 u \bar \nu \gamma_{\alpha} (1 - \gamma_5) l
 \biggl )\;,
\end{eqnarray}
where $G_f$ is the Fermi constant, $\theta_c$ is the Cabibo angle, $g_s, g_p, g_v, g_a$ are the 
scalar, pseudoscalar, vector, and pseudovector constants. From Lagrangian (\ref{Lagr}), the matrix element for  $K( p ) \to \pi^0 ( p^{'} ) l (p_l) \nu ( p_{ \nu } ) \gamma (q)$ decay can be written as follows:
\begin{eqnarray}
T = \frac {G_f} {\sqrt 2} V_{us}^* e \epsilon_{\alpha}^* \biggl (
( (1 + g_v) V^{\alpha \beta} - (1 - g_a) A^{{\alpha \beta}} ) \bar {\nu} (1+\gamma_5) \gamma_{\beta}  l )+
\label{T} \\ \nonumber
(1+g_v) F_{\beta} \bar {\nu} (1+\gamma_5) \gamma^{\beta} 
(\frac {p^{\alpha}} {p q} - \frac {p_l^{\alpha}} {p_l q} - \frac {\hat q \gamma^{\alpha}} {2 (p_l q) }) l +
\\ \nonumber
(g_s F_s^{\alpha} + g_p F_p^{\alpha}) \bar \nu (1+ \gamma_5) l + g_s f  \bar {\nu} (1+\gamma_5) 
(\frac {p^{\alpha}} {p q} - \frac {p_l^{\alpha}} {p_l q} - \frac {\hat q \gamma^{\alpha}} {2 (p_l q) }) l 
\biggr ),
\end{eqnarray}
where $\epsilon_{\alpha}$ is the photon polarization vector, and tensors $V^{\alpha \beta}, A^{\alpha \beta}, F^{\beta}, F_s^{\alpha}, 
F_p^{\alpha}, f$ can be presented in the following way: 
\begin{eqnarray}
V^{\alpha \beta} + \frac {p^{\alpha}} {p q} F^{\beta} &=& 
i \int d^4 x e^{i q x} \langle \pi^0( p^{'} )  
|T J^{\alpha} (x) (\bar s \gamma^{\alpha} u) (0) | K(p) \rangle, 
\label {tensor} \\ \nonumber
A^{\alpha \beta} &=& i \int d^4 x e^{i q x} \langle \pi^0(p^{'})  
|T J^{\alpha} (x) (\bar s \gamma^{\alpha} \gamma_5 u) (0)| K(p) \rangle,  \\ \nonumber
F_s^{\alpha} + \frac {p^{\alpha}} {p q} f &=& i \int d^4 x e^{i q x} \langle \pi^0( p^{'} )  
|T J^{\alpha} (x) (\bar s  u) (0)| K(p) \rangle, \\ \nonumber
F_p^{\alpha} &=& i \int d^4 x e^{i q x} \langle \pi^0 (p^{'}) 
|T J^{\alpha} (x) (\bar s \gamma_5 u) (0)| K(p) \rangle, \\ \nonumber
F^{\beta} &=& \langle \pi^0 (p^{'}) | (\bar s \gamma^{\beta} u) (0)| K \rangle, \\ \nonumber
f &=& \langle \pi^0 (p^{'}) | (\bar s u) (0) | K(p) \rangle\;, 
\end{eqnarray}
where $J^{\alpha}$ is the electromagnetic current. 
From the Ward identity \cite{Bijnens:1992en} we have the following relations on tensors (\ref {tensor}):
\begin{eqnarray}
q_{\alpha} V^{\alpha \beta} = 0, \\ \nonumber
q_{\alpha} A^{\alpha \beta} = 0, \\ \nonumber
q_{\alpha} F_s^{\alpha } = 0, \\ \nonumber
q_{\alpha} F_p^{\alpha } = 0.
\end{eqnarray}
Using these relations, we introduce the following parametrization of tensors:
\begin{eqnarray}
\label {formfact}
V_{\alpha \beta} = V_1 ( g_{\alpha \beta} - \frac {W_{\alpha} q_{\beta} } {q W}) + 
V_2 (p^{'}_{\alpha} q_{\beta} - \frac {p^{'} q} {q W} W_{\alpha} q_{\beta} ) + 
\\ \nonumber
V_3 (p^{'}_{\alpha} W_{\beta} - \frac {p^{'} q} {q W} W_{\alpha} W_{\beta} )+ 
V_4 (p^{'}_{\alpha} p^{'}_{\beta} - \frac {p^{'} q} {q W} W_{\alpha} p^{'}_{\beta} ), 
\\ \nonumber
A_{\alpha \beta} = i \epsilon_{\alpha \beta \rho \sigma} (A_1 p^{' \rho} q^{\sigma} +
A_2 q^{\rho} W^{\sigma}) + i \epsilon_{\alpha \lambda \rho \sigma} p^{' \lambda} q^{\rho}
W^{\sigma} ( A_3 W_{\beta} + A_4 p^{'}_{\beta}), 
\\ \nonumber
F^{\beta} = C_1 p^{'}_{\beta} + C_2 (p-p^{'})^{\beta},
\\ \nonumber
F_s^{\alpha} = S (p^{\alpha} - \frac {p q} {p^{'} q} p^{{'} \alpha}),
\\ \nonumber
F_p^{\alpha} = i P \epsilon^{ \alpha \lambda \rho \sigma} p_{\lambda} p^{'}_{\rho} q_{\sigma},
\\ \nonumber
W = p_l +p_{\nu_l},
\end{eqnarray}
We use the values of $V_i, A_i, C_i$ formfactors, calculated in the framework of $\chi$PT up to
$O(p^4)$ \cite{Bijnens:1992en}.
The values of $S, f$ formfactors can be found from $V_i, C_i$ ones using the Ward identities. 
The derivation of these relations is given in Appendix, where the value of $P$ formfactor is  calculated as well.

In searches of possible $CP$-violating effects we are interested in the distribution  of partial $K^+(p) \to \pi^0(p') l(p_l) \nu(p_{\nu}) \gamma(q)$ decay width over the kinematical variable 
$\xi={ {\bf q} \cdot [{\bf p_\pi} \times {\bf p_l}]} / {m_K^3}$ in 
the $K^+$-meson rest frame:
\begin{equation}
\rho ( \xi ) = \frac {d \Gamma} {d \xi}\;.
\end{equation}
Obviously, $\rho(\xi)$ function can be written as 
\begin{displaymath}
\rho=f_{{\mbox{\small even}}}(\xi)+f_{{\mbox{\small odd}}}(\xi)\;,
\end{displaymath}
where $f_{{\mbox{\small even}}}(\xi)$ and $f_{{\mbox{\small odd}}}(\xi)$  are even and odd functions of $\xi$, correspondingly.
$f_{{\mbox{\small odd}}}(\xi)$ can be rewritten as follows
\begin{equation}
f_{{\mbox{\small odd}}}=g(\xi^2) \xi\;.
\end{equation}
Integrating $\rho ( \xi)$ over the kinematical region, one can notice that the contribution of $f_{{\mbox{\small odd}}}( \xi)$ vanishes.

\noindent
To analize $K^+ \to \pi^0 l^+ \nu_l \gamma$ decay data we introduce the following observable:
\begin{equation}
A_\xi =\frac {N_+ - N_-} {N_+ + N_-}\;,
\end{equation}
where $N_+$ and $N_-$ are the number of events with $\xi >0$ and $\xi <0$, correspondingly. Obviously, the numerator of $A_\xi$ depends only on $f_{{\mbox{\small odd}}}( \xi )$. 

Due to the fact that $V_{i}, A_i, C_i$ formfactors are real at tree level of SM, the distribution of
$\rho ( \xi )$ is symmetrical with repect to the  $\xi =0$ line, i.e. the numbers of events with 
$\xi >0$ and $\xi < 0$ in $K^+ \to \pi^0 l^+ \nu_l \gamma$ decay are equal and $A_{\xi}=0$.
This can be easily explained:
at the SM tree level  the formfactors  $V_{i}, A_i, C_i$ are real, the matrix element squared depends only on the scalar products of momenta of $\pi, \mu , \nu, \gamma$ and the terms linear in $\xi$ vanishes. Therefore $\rho ( \xi)$ is the even function of $\xi$.
The $\xi$-odd SM terms appear due to the electromagnetic final state interaction. This leads to appearance of non-zero imaginary parts of formfactors, that, in turn, gives the nonvanishing contribution to the $f_{odd} ( \xi )$ function and  asymmetry, $A_{\xi}$.

The contribution of the one-loop final state interaction  to $A_{\xi}$ was considered earlier in \cite{Braguta:2001nz}. Calculation used $S$-matrix unitarity led to the following result:
\begin{eqnarray}
A_{\xi} = 1.14 \times 10^{-4}~~ K^+ \to \pi^0 \mu^+ \nu_{\mu} \gamma 
\label{A} \\ \nonumber
A_{\xi} = -0.59 \times 10^{-4}~~ K^+ \to \pi^0 e^+ \nu_{e} \gamma 
\end{eqnarray}
This result indicates that the SM contribution to $K^+ \to \pi^0 l^+ \nu_l \gamma$ decay asymmetry is suppressed, thus making the $A_{ \xi}$ observable quite perspective to search for a new physics.

Let us consider the value of asymmetry $A_{\xi}$ induced by Lagrangian (\ref{Lagr}). In the 
$K^+$-meson rest frame  the squared decay amplitude (\ref {T}) can be presented as 
\begin{eqnarray}
|T|^2 =	|T_{even}|^2 + (Im(g_v) C_v  + Im(g_a) C_a + Im(g_s) C_s + Im(g_p) C_p) m_K^4 \xi.
\label{T2}
\end{eqnarray}
The first term  is a $\xi$-even part of matrix element squared, and the second one is a $\xi$-odd part, 
$C_a, C_v, C_s, C_p$  are the kinematical factors, which depend only on the scalar products of final particle momenta. We do not present here the explicit expressions for $C_a, C_v, C_s, C_p$, as they seems to be quite cumbersome. 
It follows from (\ref{T2}) that the asymmentry has non-zero value only when there are non-zero imaginary parts of Lagrangian (\ref {Lagr}) parameters. 

From the expression for matrix element (\ref {T}) one can obtain the relation between $C_a, C_v$ kinematical factors. Let us suppose that we use the model with $Im(g_v)=-Im(g_a)$ and $Im(g_s)=Im(g_p)=0$. The matrix element in this model differs from SM one only 
by a total phase, that can not lead to non-zero asymmetry.
So, in this model $C_v-C_a=0$, and since the inner structure of the model does not affect the kinematical factors $C_v, C_a$, it would be correct to state that $C_v=C_a$ in any model.

Integrating the squared amplitude (\ref{T2}) over the phase space one can obtain the $A_{\xi}$ value. The value of this asymmetry, averaged over the kinematical region 
$E_\gamma >30 $~MeV è $\theta_{\gamma \l}>20^0$ in kaon rest frame, is equal to
\begin{eqnarray}
\nonumber
&&K^+ \to \pi e \nu_e \gamma: \\ \nonumber
&&A_{\xi}=-(2.9 \times 10^{-6} Im(g_s)+3.7 \times 10^{-5} Im(g_p)+ 3.0 \times 10^{-3} 
Im(g_v+g_a) )
\\ \nonumber
&&K^+ \to \pi \mu \nu_\mu \gamma: \\ \nonumber
&&A_{\xi}=-(3.6 \times 10^{-3} Im(g_s)+1.2 \times 10^{-2} Im(g_p)+ 1.0 \times 10^{-2} 
Im(g_v+g_a) )
\label{assim}
\end{eqnarray}
It should be noticed that in the formula for asymmetry of $K^+ \to \pi e \nu_e \gamma$ decay amplitude the contributions of $Im(g_s), Im(g_p)$ parameters are suppressed in comparison with the case of 
$K^+ \to \pi \mu \nu_\mu \gamma$ decay one, that is caused by the proportionality of kinematical factors multiplying these parameters to the final lepton masses.

\section{ $SU(2)_L \times SU(2)_R \times U(1)$ models}

\noindent
In this section the extensions of SM based on the $SU(2)_L \times SU(2)_R \times U(1)$ \cite{lrm} gauge group are considered. In these models each generation of fermions is formed in $SU(2)_L$ and $SU(2)_R$ doublets. At least, one Higgs multiplet $\Phi$(2,2,0) is introduced to generate the fermion masses:
\begin{eqnarray}
\Phi = 
\left (%
\begin{array} {cc}
\phi^0_1 & \phi^+_1 \\
\phi^-_2 & \phi^0_2 \\
\end{array} %
\right )\;,
\end{eqnarray}
which vacuum expectation value can be presented as follows:
\begin{eqnarray}
\Phi = 
\left (%
\begin{array} {cc}
k & 0 \\
0 & k' \\
\end{array} %
\right ).
\end{eqnarray}
Generally, the vacuum expectation values, $k, k'$, are complex. Additional Higgs multiplets are required to break $SU(2)_L \times SU(2)_R \times U(1)$ symmetry to $U(1)$. In the simpliest case two doublets, $\delta_L$(2,1,1) and $\delta_R$(1,2,1), are introduced:
\begin{eqnarray}
\delta_L =  
\left (%
\begin{array} {c}
\delta^+_L  \\
\delta^0_L  \\
\end{array} %
\right ),~~~~
\delta_R =  
\left (%
\begin{array} {c}
\delta^+_R  \\
\delta^0_R  \\
\end{array} %
\right ).
\end{eqnarray}
For a large mass scale $M_R$ it is necessary to have 
the vacuum expectation value $<\delta^0_R>=v_R$  greater than $k, k',<\delta_L^0>=v_L$.

Other scenario of spontaneous $SU(2)_L \times SU(2)_R \times U(1)$ symmetry violation is also possible. In this case two triplets $\Delta_L$(1,3,2) and  $\Delta_L$(3,1,2) are introduced
\begin{eqnarray}
\Delta_{L,R} = 
\left (%
\begin{array} {cc}
 {\Delta^+} / \sqrt 2 & \Delta^{++} \\
\Delta^0 & -  \Delta^+ / \sqrt 2 \\
\end{array} %
\right )_{L,R}\;,
\end{eqnarray}
whose vacuum expectation values can be expressed as:
\begin{eqnarray}
\Delta_{L,R} = 
\left (%
\begin{array} {cc}
 0 & 0 \\
 v_{L,R} & 0 \\
\end{array} %
\right ).
\end{eqnarray}
As in the model with two Higgs doublets the condition $v_R \gg k, k', v_L$ should be valid for large mass scale appearance.

One can also require the Lagrangian of the model under investigation to be invariant under following transformations
\begin{eqnarray}
\Psi_L \leftrightarrow \Psi_R,~ \delta_R \leftrightarrow \delta_L,
~\Delta_R \leftrightarrow \Delta_L,~ \Phi \leftrightarrow \Phi^+\;,
\label{simmetry}
\end{eqnarray}
that  leads to the fermions -- $SU(2)_L$ and $SU(2)_R$-bosons coupling constants are to be equal.

In this case, $CP$-violation appears due to the CKM matrices in left ($K^L$) and right ($K^R$)
sectors of the model. The effects of $CP$-violation are more deep  than in SM, as the analogue of the CKM matrix for right sector of the theory contains $N (N+1)/2$ phases, where $N$ is the number of fermion generations.
Notice that in our calculations we neglect the neutrino masses, that allows to neglect the lepton mixing matrices and consider them as diagonal.
Depending on the model parameters there are two possible mechanisms of $CP$-violation.
The first one is the spontaneous $CP$-violation, due to the complexity of $k, k', v_R, v_L$, where the matrix of the Yukawa couplings of $\Phi$ with fermions is real.
Second scenario assumes the complexity of Yukawa constants when the vacuum expectation values are real.
The latter mechanism applies in SM. In general case, both variants are possible.

The interaction of charged gauge bosons with quarks may be written as \cite{Langacker:1989xa}:
\begin{eqnarray}
L= \frac {g_L} {\sqrt 2} W_L^{\mu} \bar U \gamma_{\mu} K^L P_L D +
 \frac {g_R} {\sqrt 2} W_R^{\mu} \bar U \gamma_{\mu} K^R P_R D + H.C.,
\label {Llrm}
\end{eqnarray}
where $g_R, g_L$ are the coupling constants of right ang left sectors of the model, $U^T=(u, c, t)$ and  $D^T=(d, s, b)$ are the quark physical states, and $P_{L,R}=(1 \mp \gamma_5)/2$. States $W_R, W_L$ are   
non-physical states, but it is possible to make them physical performing the unitary transformation
\begin{eqnarray}
\left (%
\begin{array} {c}
W_L \\
W_R \\
\end{array} %
\right ) = 
\left (%
\begin{array} {cc}
\cos \eta  & - \sin \eta \\
e^{i \omega} \sin \eta  & e^{i \omega} \cos \eta  \\
\end{array} %
\right ) 
\left (%
\begin{array} {c}
W_1 \\
W_2 \\
\end{array} %
\right ), 
\label{Mix}
\end{eqnarray}
where $\eta$ is the mixing angle, $\omega$ is the phase. In the following calculations the phase factor will be included in $K^R$ matrix. From the formulae (\ref{Llrm}), 
(\ref {Mix}) the effective Lagrangian of the $s \to u \mu \nu_{\mu}$ process can be written as:
\begin{eqnarray}
L=- \frac {G_F} {\sqrt 2}  \frac {g_R} {g_L} K^{R*}_{su} \eta 
(\bar s (1 - \gamma_5) \gamma_{\alpha} u) (\bar \nu (1+ \gamma_5) 
\gamma^{\alpha} l).
\end{eqnarray}

Comparing the latter expression with Lagrangian in (\ref{Lagr}) one can obtain the expression for $g_a, g_v$ parameters:
\begin{eqnarray}
\nonumber
g_v = g_a= -	\frac {g_R} {g_L} \frac {K^{R*}_{su}} {sin \theta_c} \eta \\
\end{eqnarray}
Futher we suppose that the model Lagrangian is invariant under  transformation (\ref{simmetry}), that gives us the following identity: $g_R=g_L$. Moreover, we suppose that $K_R, K_L$ matrices are related as follows: $|(K_R)_{ij}|=|(K_L)_{ij}|$. This condition applies when the vacuum expectation values of Higgs fields are real, i.e. $CP$-violation appears due to complexity of the Yukawa couplings. In this case $K_L=K_R$ \cite{Masiero:1981zd}. The identity $|(K_R)_{ij}|=|(K_L)_{ij}|$ is also valid in the models with spontaneous $CP$-violation. Here, the Yukawa  matrix is real and symmetric, but vacuum expectation values are complex, that leads to $K_L=(K_R)^*$ identity \cite{Frere:1991db}. Using this relation one can rewrite $K_R$ as follows:
\begin{eqnarray}
K^R= e^{i \gamma}
\left (%
\begin{array} {cc}
e^{-i \delta_2} \cos \theta_c & e^{-i \delta_1} \sin \theta_c \\
- e^{-i \delta_1} \sin \theta_c & e^{i \delta_2} \cos \theta_c \\
\end{array} %
\right ).
\end{eqnarray}
From the explicit expression for $K^R$ matrix the imagimary parts of $g_a, g_v$ can be written as:
\begin{eqnarray}
Im ( g_a )=Im ( g_v )= - \eta \sin (\gamma - \delta_1).
\end{eqnarray}
Bounds on the model parameters, $M_R > 715 $ GeV, $\eta < 0.013$, 
have been derived from the  low-energy data
\cite{Czakon:1999ga}. 

Using these bounds and inequality $|Im (g_a)| = |Im (g_v)| < \eta$, one can derive the following upper bounds on the  $A_{\xi}$ value:
\begin{eqnarray}
\nonumber
|A_{\xi}| &<& 2.6 \times 10^{-4}, ~~~~ K^+ \to \pi^0 \mu \nu_{\mu} \gamma, \\
|A_{\xi}| &<& 0.8 \times 10^{-4} , ~~~~ K^+ \to \pi^0 e \nu_{e} \gamma
\end{eqnarray}
These values of asymmetry can be experimentally observed only if collected experimental statistics 
is about $\sim 10^7$ for $K^+ \to \pi \mu \nu_{\mu} \gamma$ decay and $\sim 10^8$ for
the $K^+ \to \pi e \nu_e \gamma$ one.

Let us now estimate the potential of T-odd correlation. Obviously, if 
T-odd correlation  coming from new physics is of order of 
it SM background (\ref{A}), the task to measure the quantity will become much 
more complicated. Making use of this fact one may obtain the parameters
of the Lagrangian under investigation when this problem arises

\begin{eqnarray}
\nonumber
|Im(g_a)|=|Im(g_v)| < 5.7 \cdot 10^{-3}, ~~~~ K^+ \to \pi^0 \mu \nu_{\mu} \gamma,  \\
|Im(g_a)|=|Im(g_v)| < 9.8 \cdot 10^{-3},~~~~  K^+ \to \pi^0 e \nu_{e} \gamma
\end{eqnarray}

One may compare this values with the bounds from \cite{Czakon:1999ga} 
$|Im (g_a)| = |Im (g_v)| < 0.013$. Although the improvement of this
restrictions is not too big, the experiment measuring the asymmetry 
could put model independent restrictions on the vector and pseudovector 
parameters. 

\section{Models with scalar interaction}

\noindent
In this section we consider the models with $Im(g_a)=Im(g_v)=0$. For this case the non-zero asymmetry apears only due to the nonvanishing values of  $Im(g_s), Im(g_p)$ parametres.
Among these models there are some leptoquark and multihiggs SM extentions \cite{Belanger:1991vx, Chen:1997gf, Cheng:42}.
  
Note, that the $K^+ \to \pi e \nu_e \gamma$  decay is not perspective
to probe such models. 
It follows from  the proportionality of kinematical factors $C_s, C_p$, entering the formula for asymmetry (\ref{assim}), to the lepton mass, that leads to the suppression of scalar and pseudoscalar contributions to this asymmetry. Moreover, in multihiggs models the additional suppression appears due to the fact that the Yukawa couplings are proportional to the fermion mass.

So, from two decays considered, only $K^+ \to \pi \mu \nu_{\mu} \gamma$ 
seems to be perspective. 
To set upper limits on possible asymmetry value in this decay let us consider the muon transverse polarization in $K^+ \to \pi \mu \nu_{\mu}$. Model independent investigation of muon transverse polarization in this decay \cite{Belanger:1991vx} allows one to claim that $P_T$ is not sensitive to $g_v, g_a, g_p$ constants. 

In order to set bounds on $Im(g_s)$ constant, one have to write down the matrix element of 
$K(p)^+ \to \pi^0(p') \mu(p_{\mu}) \nu_{\mu} (p_{\nu})$ decay:
\begin{eqnarray}
M = \frac {G_f} 2 sin \theta_c  (f_{+} (p+p')^{\lambda} +f_-  (p-p')^{\lambda})
\bar u(p_{\nu})(1+\gamma_5) \gamma_{\lambda} v(p_{\mu})\;.
\end{eqnarray}
The data of the KEK-E246 experiment on the transverse polarization
measurement gives the following result for the value of 
$Im ( \chi )= Im( f_- / f_+)$  \cite{Abe:2002vc}:
\begin{eqnarray}
Im (\chi) =(-0.28 \pm  0.69(stat.) \pm 0.30(syst.)) \times 10^{-2}
\label {imchi}
\end{eqnarray}
Using the expression for effective Lagrangian (\ref {Lagr}) one can relate the values of $Im (\chi )$ and $Im(g_s)$:
\begin{eqnarray}
Im ( \chi) = Im (g_s) \frac {m_K^2} {m_{\mu} m_s} 
\label {imgs}
\end{eqnarray}
From equations (\ref {imchi}), (\ref {imgs}) it is easy to obtain the following upper limit:   $|Im(g_s)|<6.7 \times 10^{-4}$. Futher, we assume that $Im(g_p) \sim -Im(g_s)$. This assumption is valid in any model if one neglects  the mass of $u$-quark. Obviously, within such approach it is not necessary to consider the inner structure of the models.  
Using the upper limit on the model parametes one can derive the bounds the on the $A_{\xi}$ asymmetry of the
$K^+ \to \pi^0 \mu \nu_{\mu} \gamma$ decay:
\begin{eqnarray}
|A_{\xi}|< 6.0 \times 10^{-6}.
\label {reskek}
\end{eqnarray}
Having this upper limit on the $A_{\xi}$ value one can suppose that for reliable observation of this asymmetry it is necessary to have more than $\sim 10^{10}$  events. 

If we compare values (\ref {reskek}) with SM background one may see that 
no perspective to improve restrictions to scalar and pseudoscalar parameters
are expected.

\section{Conclusion}

\noindent
The asymmetry $A_{\xi}$ in the $K^+ \to \pi l \nu \gamma$  decays is investigated in the framework of models corresponing to effective Lagrangian (\ref{Lagr}) up to $O(p^4)$ terms of $\chi$PT .

It is shown that the scalar and pseudoscalar sectors of Lagrangian contribute to asymmetry  $A_{\xi}$. However, since the kinematical factors in (\ref{T2}) are proportional to the lepton mass, $A_{\xi}$ is strongly suppresed in the $K^+ \to \pi e \nu_e \gamma$ decay.
 As for the decay with muon in the final state, the dependence of the asymmentry on scalar interaction effects is strong enough. 
KEK-E246 data allows one to obtain the strict bounds on the coupling constant, that significantly sophisticates the search for the  scalar and pseudoscalar interaction contributions 
to asymmetry . It is necessary at least $\sim 10^{10}$ events to observe this asymmetry experimentally, and the anticipated value of this observable could be as follows:
\begin{eqnarray}
|A_{\xi}|< 6.0 \times 10^{-6},
\end{eqnarray}
that two orders less than the SM contribution to $A_{\xi}$.  

KEK-E246 experiment allows to set  constraints strict enough only for the cvase of the pseudoscalar and scalar constants, while vector and pseudovector sectors of Lagrangian remains obscure.
Probably, the bounds on these parameters can be obtained in OKA experiment, that will come to operation in the nearest future. Our results reveal a high sensitivity of the asymmetry $A_{\xi}$ to vector and pseudovector interactions of the effective Lagrangian. 
Besides, in order to search for $CP$-violating effects, one can consider the $K^+_{l3\gamma}$-meson decays with electron and muon in the final state.
Taking into account the bounds on the parameters of $SU(2)_L \times SU(2)_R \times U(1)$ model, one can get the upper limit on the $A_{\xi}$ value 
\begin{eqnarray}
\nonumber
|A_{\xi}| &<& 2.6 \times 10^{-4}, ~~~~ K^+ \to \pi^0 \mu \nu_{\mu} \gamma, \\
|A_{\xi}| &<& 0.8 \times 10^{-4} , ~~~~ K^+ \to \pi^0 e \nu_{e} \gamma
\end{eqnarray}
Therefore, if statistic in OKA experiment is enlarged by one order 
it will possible to claim that the planning OKA experiment is quite 
perspective to probe vector and pseudovector sectors of Lagrangian, and asymmetry  
$A_{\xi}$ is quite effective observable for a new physics searches.

\section*{Acknowledgements}

\noindent
The work is supported, in part, by the Russian Foundation of Basic Research grant 01-02-16585,
Russian Education Ministry grant E02-3.1-96 and CRDF grant MO-011-0.
Authors would like to thank Kiselev V.V., Likhoded A.K. and Obraztsov V.F. for valuable remarks and fruitful discussions.

\newpage

\section{Appendix}

In the framework of $\chi$PT the QCD Lagrangian is introduced with external sources  \cite{Pich:1995bw}:
\begin{eqnarray}
L=L_{QCD}+ \bar q \gamma_{\mu} ( v^{\mu} + \gamma_5 a^{\mu}) q -
\bar q (s - i \gamma_5 p) q,
\label{lqcd}
\end{eqnarray}
where $L_{QCD}$ is the massless QCD Lagrangian,
$q^T=(u,d,s)$ are the quark fields, $v_{\mu}, a_{\mu}, s ,p$ are the 
Hermitian matrices $3 \times 3$. It is easy to see that Lagrangian
(\ref {lqcd}) is invariant under local transformations $SU(3)_L \times SU(3)_R$:
\begin{eqnarray}
q_L \rightarrow g_L q_L, ~~~ q_R \rightarrow g_R q_R, ~~~
s+ip \rightarrow g_R (s+ i p) g_L^+ 
\label {simetry} \\ \nonumber
l_{\mu}=g_L l_{\mu} g_L^+ + i g_L \partial_{\mu} g_L^+, ~~~ 
r_{\mu}=g_R r_{\mu} g_R^+ + i g_R \partial_{\mu} g_R^+,
\\ \nonumber 
l_{\mu}=v_{\mu}-a_{\mu}, ~~~ r_{\mu}=v_{\mu}+a_{\mu}.
\end{eqnarray}
Effective $\chi$PT Lagrangian is constructed (with symmetry (\ref {simetry}) taken into account) as the expansion in a series of external momenta:
\begin{eqnarray}
L_{eff} = L_{2}	+ L_{4} + ...\;,
\end{eqnarray}
where
$L_{2}, L_{4}$ are the effective Lagrangian terms up to $O(p^4), O(p^6)$, correspondingly. Notice that $L_{2}$ is invariant under (\ref {simetry}) transformations, 
while $L_{4}$ invariance is broken due to the chiral anomaly
\cite{Bijnens:1992en, Pich:1995bw}. Nevertheless, the effective Lagrangian is invariant under the transformation:
\begin{eqnarray}
v_{\mu} \pm a_{\mu} \rightarrow g (v_{\mu} \pm a_{\mu}) g^+ + i g \partial_{\mu} g^+,
\label {sim1} 
\\ \nonumber
s + i p \rightarrow g (s + i p) g^+
\\ \nonumber
g \in SU(3).
\end{eqnarray}
Taking into account the fact that generating functional is invariant under (\ref {sim1}) transformations
\begin{eqnarray}
Z [v', a', s', p'] = Z [v, a, s, p],
\end{eqnarray}
one can transform $g=1+ i \alpha + O(\alpha^2) \in SU(3)$, and obtain the Ward identities in $\chi$PT \cite{Bijnens:1992en}:
\begin{eqnarray}
\label {uord}
< \alpha \partial_{\mu} \frac {\delta Z} {\delta v_{\mu}}>=
i <\sum_{I=v, a, s, p} [\alpha, I] \frac {\delta Z} {\delta I}>\;,
\end{eqnarray}
where $<>$ means performing the trace operation.  

Let us consider the marix element $\langle 0|T a_{\mu}^3 (x) a_{\nu}^{4+i5} (y)
V^{4-i5}_{\alpha} (z) V_{\beta}^{em} (w) |0 \rangle$, where $a_{\mu}^3 (x), a_{\nu}^{4+i5}(y)$ are the axial currents corresponding to $\pi^0$ and $K^+$ mesons, $V^{4-i5}(z)$ is the vector current of the $\bar s \to \bar u$ transition,
$V_{\beta}^{em} (w)$ is the electromagnetic current. The divergence 
$\partial_z^{\alpha}$ of this matrix element can be obtained using the Ward identities. Therefore, one have to replace $\alpha$ in equation (\ref {uord}) by 
$\lambda^4-i \lambda^5 $ and act upon the matrix element by the operator:
\begin{eqnarray}
\hat A=\frac {\delta } { \delta a_{\mu}^3 (x)} 
\frac {\delta } { \delta a_{\nu}^{4+i5} (y)}
\frac {\delta } { \delta V_{\alpha}^{em} (z)}
\end{eqnarray}
At the point of $v_{\mu}=a_{\mu}=p=0, s=M$, where $M$ is the quark mass matrix, this expression takes form:
\begin{eqnarray}
\partial^{\alpha}_z \langle 0|T a_{\mu}^3 (x) a_{\nu}^{4+i5} (y)
V^{4-i5}_{\alpha} (z) V_{\beta}^{em} (w) |0 \rangle	=~~~ 
\label {con}
\\ \nonumber
i (m_u - m_s) \langle 0|T a_{\mu}^3 (x) a_{\nu}^{4+i5} (y)
s^{4-i5} (z) V_{\beta}^{em} (w) |0 \rangle
\\ \nonumber
+\langle 0|T a_{\mu}^3 (x) a_{\nu}^{4+i5} (y)
V^{4-i5}_{\beta} (z)  |0 \rangle \delta (w-z)
\\ \nonumber
+\langle 0|T  a_{\nu}^{4+i5} (y)
a^{4-i5}_{\mu} (z) V_{\beta}^{em} (w) |0 \rangle \delta (z-x)
\\ \nonumber
-\frac 1 2 \langle 0|T a_{\mu}^3 (x) a^3_{\nu} (z) 
V_{\beta}^{em} (w) |0 \rangle \delta( z-y)
\\ \nonumber
-\frac {\sqrt 3} 2 \langle 0|T a_{\mu}^3 (x) a_{\nu}^{8} (z)
   V_{\beta}^{em} (w) |0 \rangle \delta(z-y).
\end{eqnarray}
Futher, we use the reduction formulae to relate the vacuum matrix elements of (\ref {con}) with one of the $K^+ \to \pi^0$ transition. Obviously, the latter three terms in experssion (\ref {con}) do not contribute to the final result, since they do not contain any pole terms on $\pi^0$ and $K^+$ meson mass scale simultaneously. Using this fact one can rewrite this expression as follows:
\begin{eqnarray}
\partial_{\nu}^y \langle \pi^0 |T V_{\mu}^{em} (x) V^{4-i5}_{\nu} (y) | K^+ \rangle 
= \langle \pi^0 |V^{4- i 5}_{\nu} (y) |K^+ \rangle \delta (x-y)+  
\\ \nonumber
i (m_u - m_s) \langle \pi^0 |T V_{\mu}^{em} (x) S^{4- i 5} (y) |K^+ \rangle. 
\end{eqnarray}
The relation between scalar and vector formfactors in terms of
(\ref {tensor}) has the form:
\begin{eqnarray}
V^{\mu \nu} W_{\nu} + (F^{\mu} - \frac {F^{\nu} q_{\nu}} {pq} p^{\mu}) = (m_u -m_s) F_s^{\mu}. 
\end{eqnarray}
Similarly, one can obtain the expression for $f$:
\begin{eqnarray}
F^{\nu} (p_{\nu} - p'_{\nu}) =(m_u - m_s) f.
\end{eqnarray}
Non-zero contribution to formfactor $P$ can appear only due to the anomalous terms of effective $\chi$PT Lagrangian. It has the following form \cite{Bijnens:1992en}:
\begin{eqnarray}
L_{anom}(\Phi^3 \gamma)=-i \frac {e \sqrt {2} } {4 \pi^2 f_{\pi}^3}
\epsilon^{\mu \nu \rho \sigma} A_{\sigma} <Q \partial_{\mu} \Phi
\partial_{\nu} \Phi \partial_{\rho} \Phi>,
\end{eqnarray}
where $\Phi$ is the pseudoscalar mesons octet matrix, $Q=1/3 \times diag(2,-1,-1)$,
$f_{\pi}=93.2$ MeV.
Feynman diagram, contributing to formfactor $P$ is shown in Fig. 1.
Taking  into account this diagram  one can rewrite the expression for the formfactor in the following form
\begin{eqnarray}
P= \frac {\sqrt {2}} {4 \pi^2 f^2} \frac 1 {W^2-M_K^2} \frac {M_K^2} {m_s+m_u}, 
\end{eqnarray}
where $W=p-p'-q$.

\newpage

\section*{Figure captions}

\vspace*{1.5cm}
\noindent
\begin{description}
\item[Fig. 1.] Feynman diagram that gives the non-zero contribution to the formfactor $P$.
\end{description}

\newpage

\begin{figure}[ph]
\vspace*{14.cm}
\hspace*{0.5cm}
\begin{picture}(20,20)
\put(90,205){\epsfxsize=10cm \epsfbox{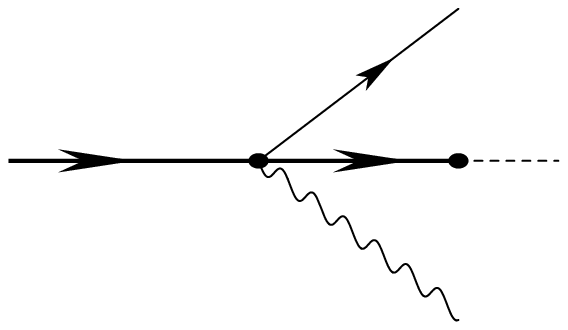}}
\put(190,150){Fig. 1}
\put(270,240){${\gamma}$}
\put(270,340){${\pi^0}$}
\put(170,300){\mbox{$K^+$}}
\put(280,300){\mbox{$K^+$}}
\put(320,300){\mbox{$H^+$}}
\end{picture}
\end{figure}

\end{document}